\documentclass[intlimits,twoside,a4paper]{article}

\usepackage{amsmath,amssymb}
\usepackage{graphicx}

\usepackage{color}
\usepackage{cancel}

\usepackage[T2A]{fontenc}
\usepackage[cp1251]{inputenc}

\usepackage{cmpj2}

\issue{2016}{19}{1}{13608}
\doinumber{10.5488/CMP.19.13608}

\title[Maier-Saupe nematogenic fluid with isotropic Yukawa repulsion at a hard wall]
{{Maier-Saupe nematogenic fluid with isotropic Yukawa repulsion at a hard wall: \\ Mean
field approximation}}

\author[M. Holovko et al.]{M. Holovko\refaddr{label1}, T. Patsahan\refaddr{label1}\thanks{E-mail: tarpa@icmp.lviv.ua}\,, I. Kravtsiv\refaddr{label1},
 D. di Caprio\refaddr{label2}
        }
\addresses{
\addr{label1} Institute for Condensed Matter Physics of the National Academy of Sciences of Ukraine, \\ 1 Svientsitskii St., 79011 Lviv, Ukraine
\addr{label2} Institut de Recherche de Chimie Paris, CNRS --- Chimie ParisTech, \\
11 rue Pierre et Marie Curie, 75005 Paris, France}

\date{Received December 31, 2015, in final form January 12, 2016}
\authorcopyright{M. Holovko, T. Patsahan, I. Kravtsiv,  D. di Caprio, 2016}

\begin{document}
\maketitle

\begin{abstract}
The mean field approximation is formulated within the framework of the density
field theory to study the properties of a Maier-Saupe nematogenic fluid
near a hard wall. The density and the order parameter
profiles are obtained using the analytical expressions derived in the linearized mean field
approximation. The temperature dependencies of the contact values of the density and
order parameter profiles are analyzed in detail. To estimate a validity of the applied
approximations, the obtained theoretical results are compared with the original computer
simulation data.



\end{abstract}

\section{Introduction}

It is a  great pleasure and a big honor for us to contribute this paper to
the festschrift dedicated to Professor Stefan Soko\l{}owski. Stefan is a
well-known expert on the modelling of physico-chemical properties of complex
fluids such as chemically reacting fluids in  disordered porous media
\cite{Trokh96,Trokh97}, studies of the isotropic-nematic phase transition in
confined nematic fluids \cite{IlnSok99}, modelling of the properties of fluids
in pores with walls decorated with tethered polymer brushes \cite{IlnPat11},
treatment of phase behavior in confined functional colloids \cite{SokKal2014}
and many other complex fluids in complex confinements.

In this paper, we study the influence of the surface on a nematic fluid near
a hard wall. Due to orientational ordering, such systems have many unique
properties which are very important in the display industry.
The anchoring phenomena are among them, according to which the surface
induces a specific orientation of the nematic director with respect to the
surface \cite{Jerom91}. In order to understand the connection between
the anchoring phenomena and the interaction between nematic molecules and the
surface, there the Henderson-Abraham-Barker (HAB) approach
previously developed for isotropic fluids at the wall \cite{Hend76} has been employed \cite{SokSok04, SokSok05}. In this approach{,} the distribution of the fluid
near the wall is described by the wall-particle Ornstein-Zernike (OZ)
equation with the fluid distribution function in the bulk calculated in the
mean spherical approximation (MSA) \cite{HolSok99}. {Using this approach, it
is possible to evaluate} the role of an orientation-dependent interaction of
nematic molecules with the surface in the anchoring phenomena. However, in the
MSA, the HAB approach does not  correctly take into account the contribution from
long-range intermolecular interactions. As a result, it does not
satisfy an exact relation known as the contact theorem, which was formulated
in \cite{HendBlum79,HolBad05} for isotropic fluids near a wall, and recently
in \cite{Hol15} it was reformulated for anisotropic fluids near a hard wall.
According to this theorem, the contact value of the density profile of a
fluid  near a hard wall is determined by the pressure of the fluid in the
bulk. In \cite{Hol15}, the contact theorem was also formulated for the order
parameter profile.

An alternative way of describing fluids at a hard wall was developed within the
framework of the density field theory. In this theory, the contributions
from the mean field and from fluctuations are separated. The theory was
successfully applied to ionic fluids at a hard wall
\cite{DdiCaprio03,DdiCaprio98,DdiCaprio05,DdiCaprio07} and to simple fluids
with Yukawa-type interactions near a hard wall \cite{DdiCaprio11,KravPat15}.
It was shown that the mean field treatment of a Yukawa fluid at a hard wall
reduces to the solution of a non-linear differential equation for the density
profile, while the treatment of fluctuations reduces to the OZ equation with
the Riemann boundary condition \cite{Gahov}. The density field theory was
applied to the description of bulk properties of a nematic fluid in
\cite{Hol11,KravHol13}. The application of the density field theory to the
description of a nematic fluid at a hard wall was initiated in~\cite{HolKrav13}.

In this paper, we use the mean field approximation {to investigate the
effect} of a hard wall on the properties of a nematic fluid. We demonstrate
the principal difference between {the behavior} of the order parameter
profile obtained in the mean field approximation and its linearized version.
In order to check {the} validity of both approaches we compare the obtained
theoretical results {with computer simulations data}.

\section{Theory}

In this paper{,} we consider Maier-Saupe (MS) nematogenic fluid model
\cite{Maier60-a,Maier60-b} as one of the simplest models that {accounts} for the
isotropic-nematic phase transition. For simplification{,} we consider a fluid
of point uniaxial nematogens interacting through the pair potential
\begin{align}
\nu(r_{12},\Omega_1\Omega_2)=\nu_0(r_{12})+\nu_2(r_{12})P_2(\cos\theta_{12}), \label{potential}
\end{align}
where the first term
$\nu_0(r_{12})=\left({A_0}/{r_{12}}\right)\exp\left({-\alpha_0
r_{12}}\right)$ describes isotropic repulsion and the second term with
$\nu_2(r_{12})=\left({A_2}/{r_{12}}\right)\,\exp\left({-\alpha_2
r_{12}}\right)$ describes anisotropic attraction between particles ($A_0>0$,
$A_2<0$), $r_{12}$ denotes the distance between particles $1$ and $2$,
$\Omega=\left(\theta,\phi\right )$ are orientations of particles,
$P_2(\cos{\theta_{12}})=(3\cos^2\theta_{12}-1)/2$ is the second order
Legendre polynomial of the relative orientation $\theta_{12}$.

{It is necessary, in numerical calculations, to cut-off the potential
$\nu(r_{12},\Omega_{1}\Omega_{2})$ at some finite distance} and due to this
in expression (\ref{potential}), $\nu_{0}(r)$ and $\nu_{2}(r)$ are replaced by
$\tilde{\nu}_{i}(r)=\nu_{i}(r)$ for  $r\leqslant r_\text{c}$ and $\tilde{\nu}_{i}(r)=0$
for $r>r_\text{c}$, where $i=0,\,2$ and $r_\text{c}$ is the cut-off radius.

Within the field-theoretical formalism, the Hamiltonian is a functional of
the density field and can be written as a sum of the entropic and the
interaction terms
    \begin{align}
    \beta H[\rho(\mathbf{r},\Omega)]&=\int\rho(\mathbf{r},\Omega)\left\{\ln\left[\rho(\mathbf{r},\Omega)
    \Lambda_\text{R}\Lambda_\text{T}^3\right]-1\right\}\rd{\mathbf{r}\rd\Omega}\nonumber\\
    &+\frac{\beta}{2}\int{\nu(r_{12},\Omega_1\Omega_2)\rho(\mathbf{r}_1,\Omega_1)\rho(\mathbf{r}_2,\Omega_2)}
    \rd{\mathbf{r}_1}\rd{\mathbf{r}_2\rd{\Omega_1}\rd{\Omega_2}},
    \end{align}
where $\beta=1/k_\text{B}T$ is the inverse temperature, $\rd\Omega=(1/4\pi)\sin\theta
\rd\theta \rd\phi$ is the normalized angle element, $\rho(\mathbf{r},\Omega)$ is
particle density per angle such that
$\int{\rho(\mathbf{r},\Omega)\rd{\Omega}}=\rho(\mathbf{r})$, $\Lambda_\text{T}$ is the
thermal de Broglie wavelength of the molecules, the quantity $\Lambda^{-1}_\text{R}$
is the rotational partition function for a single molecule \cite{Gray84}.

\subsection{Mean field approximation}

In this paper{,} we restrict our consideration to the mean field (MF)
approximation  which is the lowest order approximation for the partition
function. In the canonical formalism{,} it corresponds to fixing the Lagrange
parameter $\lambda$ such that the following relation is true for the singlet
distribution function
\begin{equation}
\label{MFA1} \frac{\delta\beta H[\rho(\mathbf{r},\Omega)]}{\delta\rho(\mathbf{r},\Omega)}\bigg\vert_{\rho^\text{MF}}=\lambda.
\end{equation}
As a result, we have
\begin{align}
\label{MFA_s} \rho(\mathbf{r}_1,\Omega_1)=\rho^\text{bulk}(\Omega_1)
\exp\left\{-\beta\int\nu(r_{12},\Omega_1\Omega_2)\left[\rho(\mathbf{r}_2,\Omega_2)-
\rho^\text{bulk}(\Omega_2)\right]\rd\mathbf{r}_2\rd{\Omega_2}\right\},
\end{align}
where
\begin{align}
\label{singlet_bulk}
\rho^\text{bulk}(\Omega)&=\rho_\text{b}\frac{\exp\left[-(\kappa_2^2S_\text{b}/\alpha_2^2)\,P_2(\cos\theta)\right]}{\int\limits_0^1
\rd\cos\theta\exp\left[-(\kappa_2^2S_\text{b}/\alpha_2^2)\,P_2(\cos\theta)\right]}
\end{align}
is the singlet distribution function for the bulk nematic fluid in the MF approximation, defined
within the framework of the Maier-Saupe theory \cite{Maier60-a,Maier60-b},
$\kappa_2^2=4\pi\rho_\text{b}\beta A_2$, $\rho_\text{b}$ is the bulk value of the fluid
density, $S_\text{b}=(1/\rho_\text{b})\int\nolimits_0^1
P_2(\cos\theta)\rho^\text{bulk}(\Omega)\rd\cos\theta$ is the bulk value of the
orientational order parameter.

After integration with respect to orientation $\Omega_{2}${,} we obtain
\begin{align}
\label{MFA_s2}
\frac{\rho(\mathbf{r}_1,\Omega_{1n},\Omega_{wn})}{\rho^\text{bulk}(\Omega_1)}
=\exp\left\{-\left[V_0(\mathbf{r}_1,\Omega_{wn})-V_0^\text{b}\right]
-\frac{1}{\sqrt{5}} \sum\limits_m
Y_{2m}(\Omega_{1n})\left[V_{2m}(\mathbf{r}_1,\Omega_{wn})-V_{2m}^\text{b}\right]\right\},
\end{align}
where $\Omega_{wn}$ denotes the angle between the nematic director and the
surface, and the mean field potentials
\begin{align}
V_0(\mathbf{r}_1,\Omega_{wn})&=\beta\int\nu_0(r_{12})
\rho(\mathbf{r}_2,\Omega_{wn})\rd\mathbf{r}_2,\\
V_{2m}(\mathbf{r}_1,\Omega_{wn})&=\beta\int\nu_2(r_{12})
S_{2m}(\mathbf{r}_2,\Omega_{wn})\rd\mathbf{r}_2.
\end{align}
The bulk values of these potentials are $V_0^\text{b}=\kappa_0^2/\alpha_0^2$,
 $V_{20}^\text{b}=\kappa_2^2 S_\text{b}/\alpha_2^2$,  $V_{2m}^\text{b}=0$
for $m\ne0$, where  $\kappa_0^2=4\pi\rho_\text{b}\beta A_0$,
\begin{align}
\rho(\mathbf{r},\Omega_{wn})=\int\rho(\mathbf{r},\Omega_{1n},\Omega_{wn})\rd\Omega_{1n}
\label{hol:eq1}
\end{align}
is the density profile. The {quantities}
\begin{align}
\label{shva} S_{2m}(\mathbf{r},\Omega_{wn})=\frac{1}{\sqrt{5}}
\int\rho(\mathbf{r},\Omega_{1n},\Omega_{wn})Y_{2m}(\Omega_{1n})\rd\Omega_{1n}
=\rho(\mathbf{r},\Omega_{wn})\,S_{2m}^*(\mathbf{r},\Omega_{wn}),
\end{align}
where $S_{2m}^*(\mathbf{r},\Omega_{wn})$ are the order parameter profiles.
Far from the wall we have $S_{20}^{*}(\mathbf{r},\Omega_{wn})\rightarrow
S_\text{b}$, \linebreak $S_{2m}^*(\mathbf{r},\Omega_{wn})\rightarrow 0$ for $m\ne0$.
Simple calculations show that{,} in order to  take into account the cut-off
radius $r_\text{c}${,} one should substitute the quantities $\kappa^{2}_{i}$ by
$\tilde{\kappa}^{2}_{i}$ such that
\begin{align}
\tilde{\kappa}^{2}_{i}=4\pi\rho\beta\int_{0}^{r_\text{c}}\nu_{i}(r)r^{2}{\rd}r=
{\kappa}^{2}_{i}\left[1-\exp(-\alpha_{i}r_\text{c})-\alpha_{i}r_\text{c}\exp(-\alpha_{i}r_\text{c})\right].
\label{hol:eq2}
\end{align}

\subsection{Linearized MF approximation}

The gradient of equation~(\ref{MFA_s2}) gives
\begin{align}
\label{grad} \frac{1}{\rho(\mathbf{r},\Omega_{1n},\Omega_{wn})}\pmb\nabla\rho(\mathbf{r},\Omega_{1n},\Omega_{wn})=
\mathbf{E}_0(\mathbf{r},\Omega_{wn})+\frac{1}{\sqrt{5}}\sum\limits_m Y_{2m}(\Omega_{1n}) \mathbf{E}_{2m}(\mathbf{r},\Omega_{wn}),
\end{align}
where we define an equivalent of the electric field as
\begin{align}
\label{electric} \mathbf{E}_0({\mathbf{r},\Omega_{wn}})\equiv -\pmb\nabla V_0(\mathbf{r},\Omega_{wn}),\qquad
\mathbf{E}_{2m}({\mathbf{r},\Omega_{wn}})\equiv -\pmb\nabla V_{2m}(\mathbf{r},\Omega_{wn}).
\end{align}
According to the properties of the Yukawa potential we can write
\begin{align}
\label{hered}
\left(\triangle-\alpha_0^2\right)V_0(\mathbf{r},\Omega_{wn})&=-4\pi\beta A_0 \rho(\mathbf{r},\Omega_{wn}),\\
\label{hered1} \left(\triangle-\alpha_2^2\right)V_{2m}(\mathbf{r},\Omega_{wn})&=-4\pi\beta A_2S_{2m}(\mathbf{r},\Omega_{wn}).
\end{align}
Due to translational invariance parallel to the wall, the functions considered
 depend only on the distance $z$ to the wall. Equations
(\ref{shva})--(\ref{hered1}) {make} a set of six differential equations for
the unknown functions $\rho(\mathbf{r},\Omega_{1n},\Omega_{wn})$,
$S_{2m}(\mathbf{r},\Omega_{wn})$, $E_0(\mathbf{r},\Omega_{wn})$,
$E_{2m}(\mathbf{r},\Omega_{wn})$, $V_0(\mathbf{r},\Omega_{wn})$,
$V_{2m}(\mathbf{r},\Omega_{wn})$. We note that in the case when the director
is oriented perpendicularly to the wall, $\Omega_{wn}=0$, the singlet
distribution function is axially symmetric. Consequently, the equations
considered will retain only the terms with $m=0$. In this paper, we will
restrict our further investigation to this special case.

As was shown in~\cite{HolKrav13}{,} the differential equations obtained can
be solved analytically in the linear approximation for the expression
(\ref{MFA_s2})
\begin{align}
\rho{'}(z,\Omega)=\left[E_0(z)+E_{20}(z)P_2(\cos\theta)\right]\rho^\text{bulk}(\Omega),
\end{align}
where the prime denotes derivative by $z$.

The resulting solutions of the linearized profile are as follows:
\begin{eqnarray}
\label{dprofile} \frac{\rho(z)}{\rho_\text{b}}&=&1-\frac{\lambda_0^2-\alpha_2^2-\frac{1}{5}\kappa_2^2 \big(\,\langle Y_{20}^2\rangle_{\Omega}-
\,\langle Y_{20}\rangle_{\Omega}^2\big)}{\kappa_2^2\,S_\text{b}}\,B_1\,\re^{\displaystyle-\lambda_0 z}\nonumber \\
&&-\frac{\lambda_2^2-\alpha_2^2-\frac{1}{5}\kappa_2^2 \big(\,\langle Y_{20}^2\rangle_{\Omega}-
\,\langle Y_{20}\rangle_{\Omega}^2\big)}{\kappa_2^2\,S_\text{b}}\,B_2\,\re^{\displaystyle-\lambda_2 z},\\
\frac{S_{20}(z)}{\rho_\text{b}\,S_\text{b}}&=&1-\frac{\left(\lambda_0^2-\alpha_2^2\right)}
{\kappa_2^2\,S_\text{b}}\,B_1\,\re^{\displaystyle-\lambda_0 z}
-\frac{\left(\lambda_2^2-\alpha_2^2\right)}{\kappa_2^2\,S_\text{b}}\,B_2\,\re^{\displaystyle-\lambda_2 z}, \label{sprofile}
\end{eqnarray}
where
\begin{align}
&B_1=\frac{\kappa_2^2\,S_\text{b}}{2\left(\lambda_0^2-\lambda_2^2\right)}
\left[-\frac{\kappa_0^2}{\alpha_0^2}
+\frac{\lambda_2^2-\alpha_2^2-({\kappa_2^2}/{5})\langle Y_{20}^2\rangle_{\Omega}}{\alpha_2^2}\right],\qquad
B_2=-\frac{\kappa_2^2\,S_\text{b}}{2\alpha_2^2}-B_1,
\label{hol:eq3}\\
&\langle Y_{20}^{k}\rangle_{\Omega}=(1/\rho_\text{b})\int\limits_0^1
Y_{20}^{k}(\Omega)\rho^\text{bulk}(\Omega)\rd\cos\theta.
\end{align}
\noindent
Parameters $\lambda_{0}$ and $\lambda_{2}$
\begin{align}
\label{lambda0} \lambda_{0,2}^2=\frac{1}{2}\left\{\kappa_0^2+\alpha_0^2 +\kappa_2^2\langle P_2^2(\cos\theta)\rangle+
\alpha_2^2\pm\sqrt{\left[\kappa_0^2+ \alpha_0^2-\kappa_2^2\langle P_2^2(\cos\theta)\rangle-\alpha_2^2\right]^2
+4\kappa_0^2\kappa_2^2\,S_\text{b}^2}\right\}
\end{align}
are identical to the parameters found in the bulk phase when Gaussian
fluctuations are taken into account~\cite{KravHol13} and characterize a decay
of the isotropic repulsive and the anisotropic attractive
interactions, respectively.

Hereafter, the approach presented in this subsection is referred to as the linearized
mean field (LMF) approximation. The expressions for $\rho(z)$ and $S^*_{20}(z)$ obtained
within this approximation correspond to the case of infinite cut-off radius $r_\text{c}\rightarrow\infty$.

\subsection{Contact theorem}

As it was shown in~\cite{Hol15}{,} the density and order parameter profiles
satisfy some exact relations known as the contact theorems. According to
these relations in the absence of wall-particle interactions, the contact
values of the density profile $\rho(z=0)$ and of the order parameter profile
$S_{20}(z=0)$ do not depend on the angle $\Omega_{wn}$ and are equal to
\begin{align}\label{eq:CT2b}
  \rho(z=0) &= \beta\int \rd\Omega_{1n} P(\Omega_{1n})=\beta P,\\
  \label{eq:CTS1}
S_{20} (z=0)& = \beta \int \rd\Omega_{1n} P_{2}(\cos\theta) P(\Omega_{1n}),
\end{align}
where $P$ is the bulk pressure and $P(\Omega_{1n})$ can be treated as the
bulk partial pressure for molecules with a given orientation $\Omega_{1n}$.

In the MF approximation for the model under consideration,
relations~(\ref{eq:CT2b})~and~(\ref{eq:CTS1})~give
\begin{align}
\label{contactd}
\frac{{\rho}(0^+)}{\rho_\text{b}}&=1+\frac{\kappa_0^2}{2\alpha_0^2}+\frac{\kappa_2^2}{2\alpha_2^2}\,S_\text{b}^2{,}\\
\label{contacts}
\frac{S_{20}(0^+)}{S_\text{b}\rho_\text{b}}&=1+\frac{\kappa_0^2}{2\alpha_0^2}+\frac{\kappa_2^2}{2\alpha_2^2}\,\langle P_{2}^2(\cos\theta)\rangle.
\end{align}
These relations are used as boundary conditions in the solution of
differential equations of the LMF approximation. In order to take into
account the cutoff distance $r_\text{c}$ in
relations~(\ref{contactd})--(\ref{contacts}){,} we should change
$\kappa_{i}^{2}$ to $\tilde{\kappa}_{i}^{2}$ given by equation~(\ref{hol:eq2}).
However, we should note the principal difference between the exact results
(\ref{eq:CT2b})--(\ref{eq:CTS1}) and results (\ref{contactd})--(\ref{contacts})
of the MF approximation. In paper~\cite{Hol11}{,} an invariant
\begin{align}
\frac{\alpha_{0}^{2}}{2\kappa_{0}^{2}}V_{0}^{2}(z,\Omega_{wn})-\frac{1}{2\kappa_{0}^{2}}E_{0}^{2}(z,\Omega_{wn})
+\sum_{m}\left[\frac{\alpha_{2}^{2}}{2\kappa_{2}^{2}}V_{2m}^{2}(z,\Omega_{wn})-\frac{1}{2\kappa_{2}^{2}}E_{2m}^{2}(z,\Omega_{wn})\right]
\label{hol:eq5}
\end{align}
was found which was used to prove the contact theorem for the MF density
profile of a nematogenic fluid in form~(\ref{contactd}). However, no similar
proof of the contact theorem for the MF order parameter profile in
form~(\ref{contacts}) exists.

\section{Numerical calculation details}

In order to verify the theoretical approaches presented in the previous
section{,} a series of numerical calculations were carried out. To this
end{,} for the model pair potential (\ref{potential}), the following
parameters were chosen: $A_0/|A_2|=3.0$ and $\alpha_0/\alpha_2=1.6$. It
should be noted that all quantities marked by a star in our paper are
considered as non-dimensional. For instance, all distances are reduced as
$r^*=r\alpha_2$ or $z^*=z\alpha_2$, densities are reduced as $\rho^*=\rho
/\alpha_2^3$ and temperature as $T^*=k_\text{B}T/(|A_2| \alpha_2)$. The potential
(\ref{potential}) is characterized by a rather soft repulsive part (isotropic
contribution) and a small attractive {part} dependent on {the} relative
orientation between a pair of fluid particles (anisotropic contribution). It
is worth noting that the considered fluid is mostly a repulsive one, and a small
attractive contribution affects mainly an orientational properties of {the}
fluid, at least at the conditions used in our study. In particular{,} we
consider the fluid at the density $\rho^*_\text{b}=1.0$, and the temperature interval
$T^*=0.5-3.5$ is chosen. We have found that{,} at these temperatures{,} the
considered fluid {is beyond its vapour-liquid phase transition region.}
Therefore, only a nematic-isotropic phase transition can be expected in our case.

To obtain a numerical solution of the integral equation (\ref{MFA_s2}) used
in the MF approximation, the Picard iterative method was applied. The problem
was considered in the cylindrical coordinates, which are set along $z$-axis
{normal} to the wall surface. The integrations over $z$ were performed using
trapezoidal rule with a step $\Delta z=0.02/\alpha_2$, while all integrations
over $r$ were done analytically. A step of integration over $\cos(\theta)$
was chosen as $0.0025$. To take into account the confinement, we consider a
fluid between two hard walls at a distance $L_z=36/\alpha_2$ to each other.
Two cut-off radii $r_\text{c}=6.0/\alpha_2$ and $12.0/\alpha_2$ are used in our
study. The chosen distance $L_z$ appears to be sufficient to get a bulk-like
region of a fluid in the middle between {the two walls.} The presence of the
hard walls is {introduced} by the boundary conditions $\rho(z)=0$ if $z<0$ or
$z>L_z$. {Equation (\ref{MFA_s2}) is solved in combination with equation
(\ref{shva}) leading to density $\rho(z)$ and order parameter $S^*_{20}(z)$
profiles.} The precision of this solution expressed in terms of standard
deviation is $10^{-5}$.

The bulk order parameter $S_\text{b}$ is used both in the MF and LMF calculations.
To obtain this quantity, the integral equation (\ref{singlet_bulk}) was
applied. It was also solved numerically by the iterative method, but {here
with a} precision $10^{-12}$ and with a step of integration over
$\cos(\theta)$ taken equal to $0.00125$. The cut-off radii $r_\text{c}=6.0/\alpha_2$
and $12.0/\alpha_2$ were used to obtain $S_\text{b}$ as well.

We compare the numerical results calculated from MF and LMF approaches with Monte-Carlo (MC) simulation results.
For this purpose, a series of MC simulations in canonical ensemble~\cite{Frenkel} were performed to obtain $S_\text{b}$, $\rho(z)$ and $S^*_{20}(z)$.
A system of $N_\text{p}$ fluid particles interacting with the pair potential (\ref{potential}) were placed into a rectangular box of a size $L_x \times L_y
\times L_z$,
where $L_x=L_y=24/\alpha_2$ and $L_z=4 r_\text{c}$ if $r_\text{c}=6.0/\alpha_2$ and $L_z=3 r_\text{c}$ if $r_\text{c}=12.0/\alpha_2$.
For $L_z$ taken in our simulations, a bulk-like region in the middle of the box is observable up to the temperature $T^*=2.2$.
Since the bulk fluid density is set to $\rho^*_\text{b}=1.0$, a number of fluid particles $N_\text{p}$ is set to $13824$ for the case of $r_\text{c}=6.0/\alpha_2$
and $20736$ for the case of $r_\text{c}=12.0/\alpha_2$. The simulations were carried out with the periodical boundary conditions applied
in three dimensions in the case of a bulk fluid, and in $X$ and $Y$ dimensions in the case of a fluid between two hard walls.
Each simulation procedure was performed at a constant temperature and volume and consisted of three stages:
1)~equilibration of a fluid in a strong field applied along $Z$-axis to give the fluid particles a preferential orientation, which was normal to the wall
surfaces;
2)~equilibration of the system obtained at the previous stage with the field switched off;
3)~production of the necessary characteristics for a system obtained at the second stage with the field switched off.
A criterium defining an equilibrated system is a stabilization of the total order parameter in the system, i.e.,
when the order parameter fluctuates around one average value during an essential number of MC steps, usually it was over 20\,000 steps at least.
It should be noted that in our simulations, one MC step corresponds to $N_\text{p}$ trial translational or rotational movements.

\section{Results and discussion}

\subsection{Bulk fluid}
\begin{figure}[!h]
\begin{center}
\includegraphics [width=0.49\textwidth]{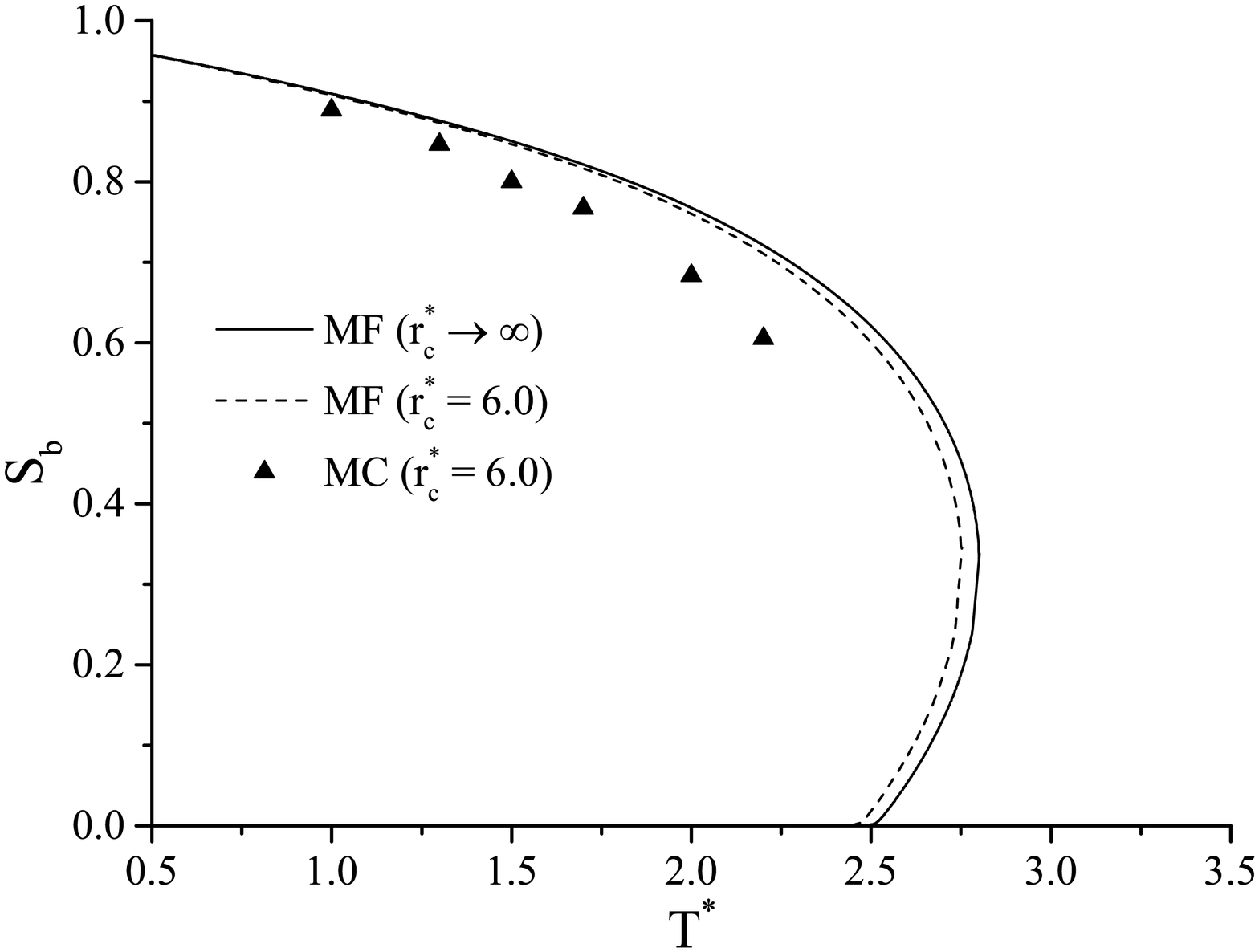}
\includegraphics [width=0.49\textwidth]{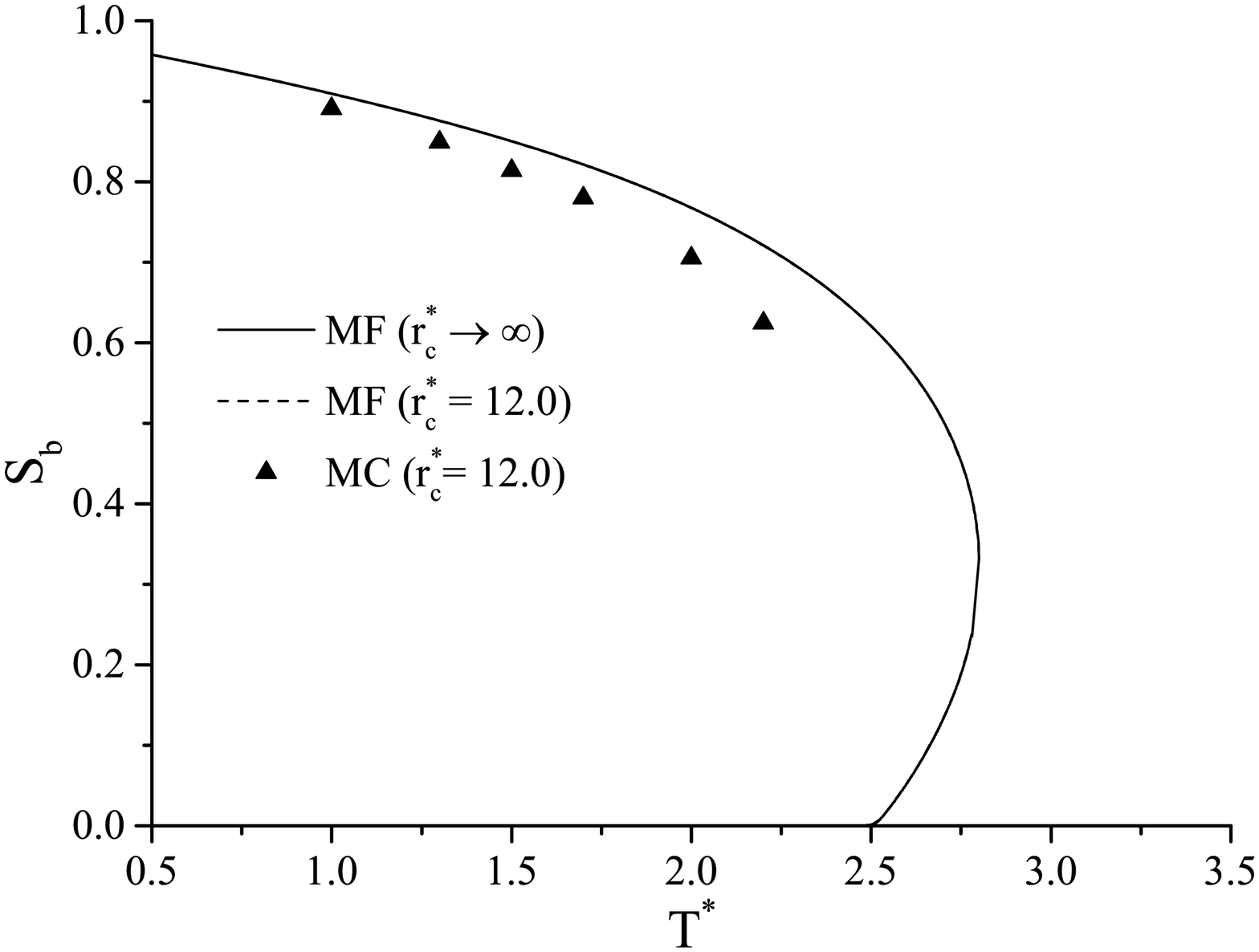}
\caption{Temperature dependence of the order parameter $S_\text{b}$ of a bulk MS fluid.
The results obtained in the MF approximation as well as
with the use of the Monte-Carlo (MC) simulation method.}
\label{fig:Fig1}
\end{center}
\end{figure}

The order parameter of the bulk MS fluid is calculated at temperatures in the
range of $T^*=0.5-3.5$ using the MF approximation with the different
cut-off radii (figure~\ref{fig:Fig1}). We have checked at which $r^*_\text{c}$ the
results tend to the case of $r^*_\text{c}\rightarrow\infty$ and have tested what  the effect of the cut-off radius in general is. This information is valuable to
make a comparison with the computer simulations in which the cut-off radius
is used. As can be seen in the case of $r^*_\text{c}=6.0$, the MF approximation
leads to the values of the order parameter lower  than in the cases of
$r^*_\text{c}=12.0$ and $r^*_\text{c}\rightarrow\infty$ for the whole stable nematic region
up to the critical region of the nematic-isotropic phase transition, which
appears at $S_\text{b}<0.443$ \cite{KravHol13} (figure~\ref{fig:Fig1}, left-hand panel).
Also, it is observed that the results for $r^*_\text{c}=12.0$ totally coincide with
the case of $r^*_\text{c}\rightarrow\infty$ (figure~\ref{fig:Fig1}, right-hand panel).
The values of the order parameter obtained from the simulations are
systematically lower than in the MF approximations (figure~\ref{fig:Fig1},
triangle symbols).

\subsection{Density and order parameter profiles}

\begin{figure}[!t]
\begin{center}
\includegraphics [width=0.49\textwidth] {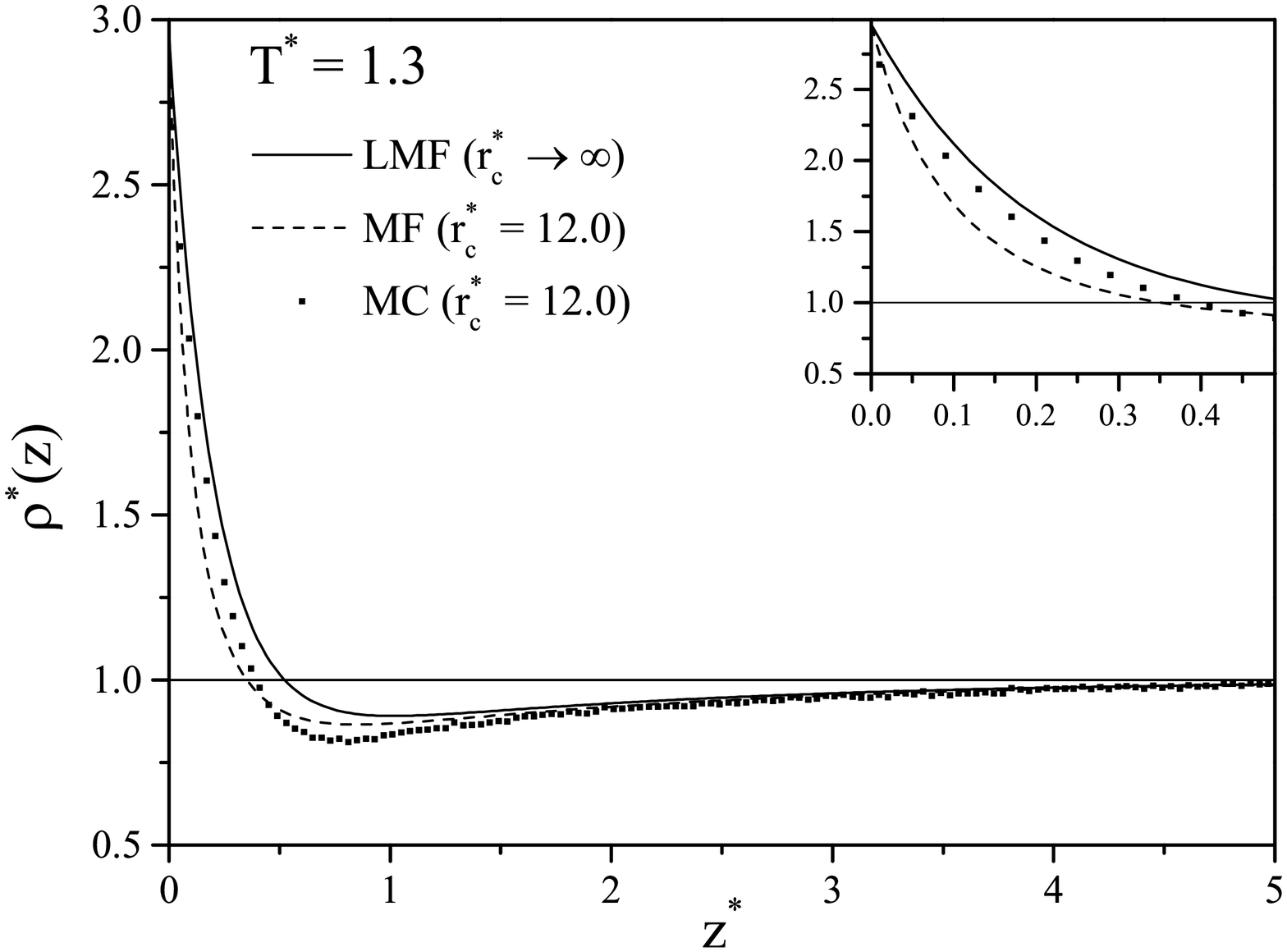}
\includegraphics [width=0.49\textwidth] {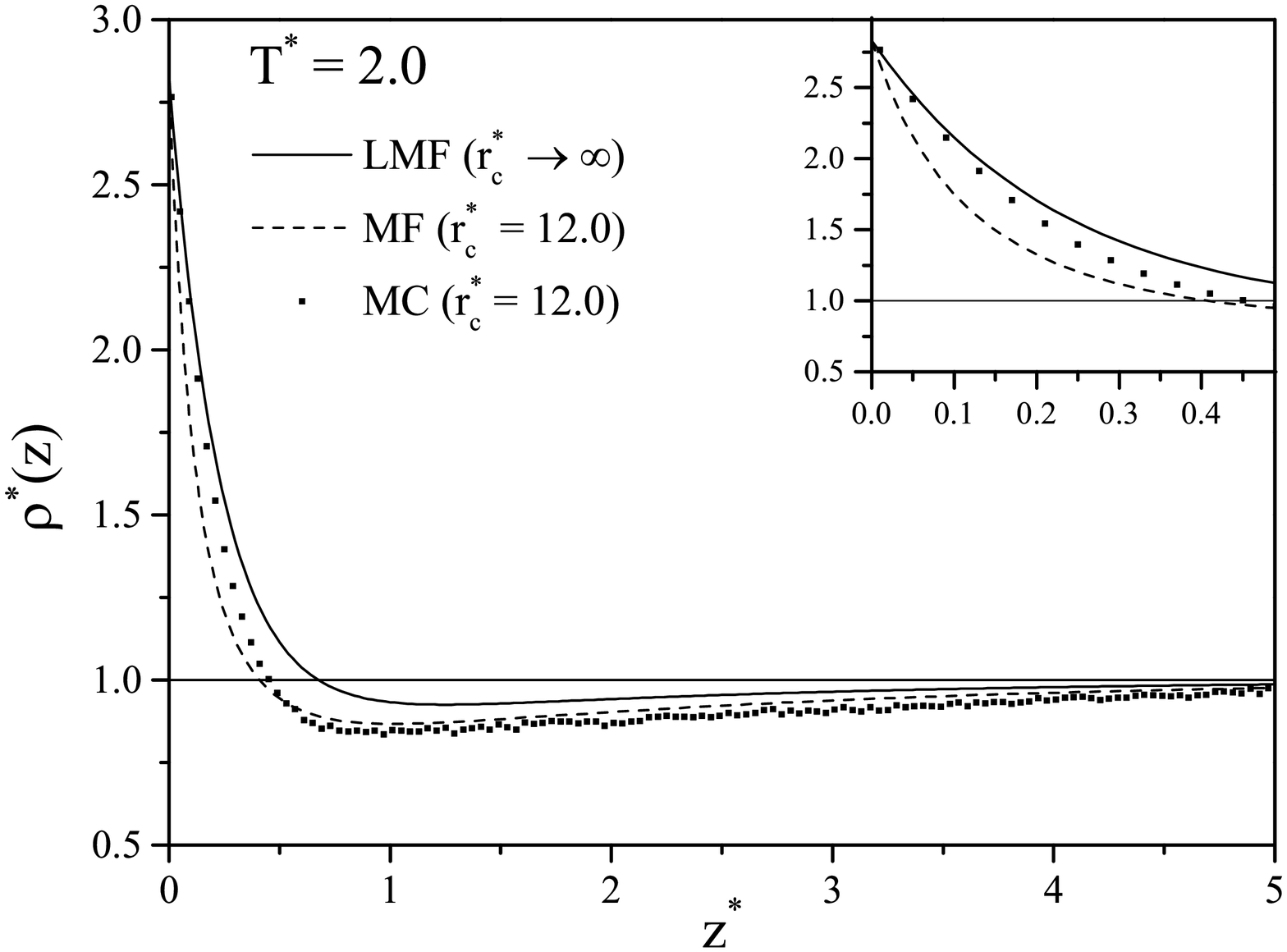}
\caption{Density profiles of the MS fluid near a hard wall obtained in the LMF and MF approximations as well as
with the use of the Monte-Carlo (MC) simulation method. The results were obtained at temperature $T^*=1.3$ (left-hand panel) and $T^*=2.0$ (right-hand panel).}
\label{fig:Fig2}
\end{center}
\end{figure}

A series of density profiles are calculated for the MS fluid near a hard wall
at different temperatures in the range $T^*=0.5-3.5$. For this purpose, the
LMF and MF approximations are applied and compared with the corresponding
simulation results. In figure~\ref{fig:Fig2}{,} we present two selected
results at temperatures $T^*=1.3$ and $2.0$. The cut-off radius $r^*_\text{c}=12.0$
is used in the MF approximation and in the computer simulations. As has been
shown above, this cut-off should be long enough to give results comparable to
the {case} $r^*_\text{c}\rightarrow\infty$. It is observed that {all presented}
profiles have the same qualitative behavior, i.e., they have a distinct high
maximum at the contact with the wall, then a minimum appears around
$z^*=0.5-1.0$ and at $z^*>5$ one can see a convergence of $\rho^*(z)$ to
$\rho^*_\text{b}$ when the bulk-like region is reached.
We note that the contact values obtained in the MF and LMF practically
coincide~--- the cut off correction is negligible, while for $\rho^*(0)$ obtained from the simulations, we observe some small
difference. The contact value of the obtained density profiles will be
discussed  more in detail later on. For this moment, we focus on $\rho^*(z)$ at
distances $z^*>0$ and on comparison of the theoretical approaches with the
computer simulations. One can see that at small distances $z^*$, the LMF
approximation agrees with the simulations better than the MF, while the MF
approximation describes $\rho^*(z)$ better at distances $z^*$ around the
minimum of $\rho^*(z)$ and larger. Moreover, at all temperatures considered in
our study, the $z^*$-position of the minimum of $\rho^*(z)$ is very close in the MF
approximation to that in the computer simulations, while in the LMF approximation,
the $z^*$-position of the minimum is notably shifted towards the higher values.

\begin{figure}[!b]
\begin{center}
\includegraphics [width=0.49\textwidth]  {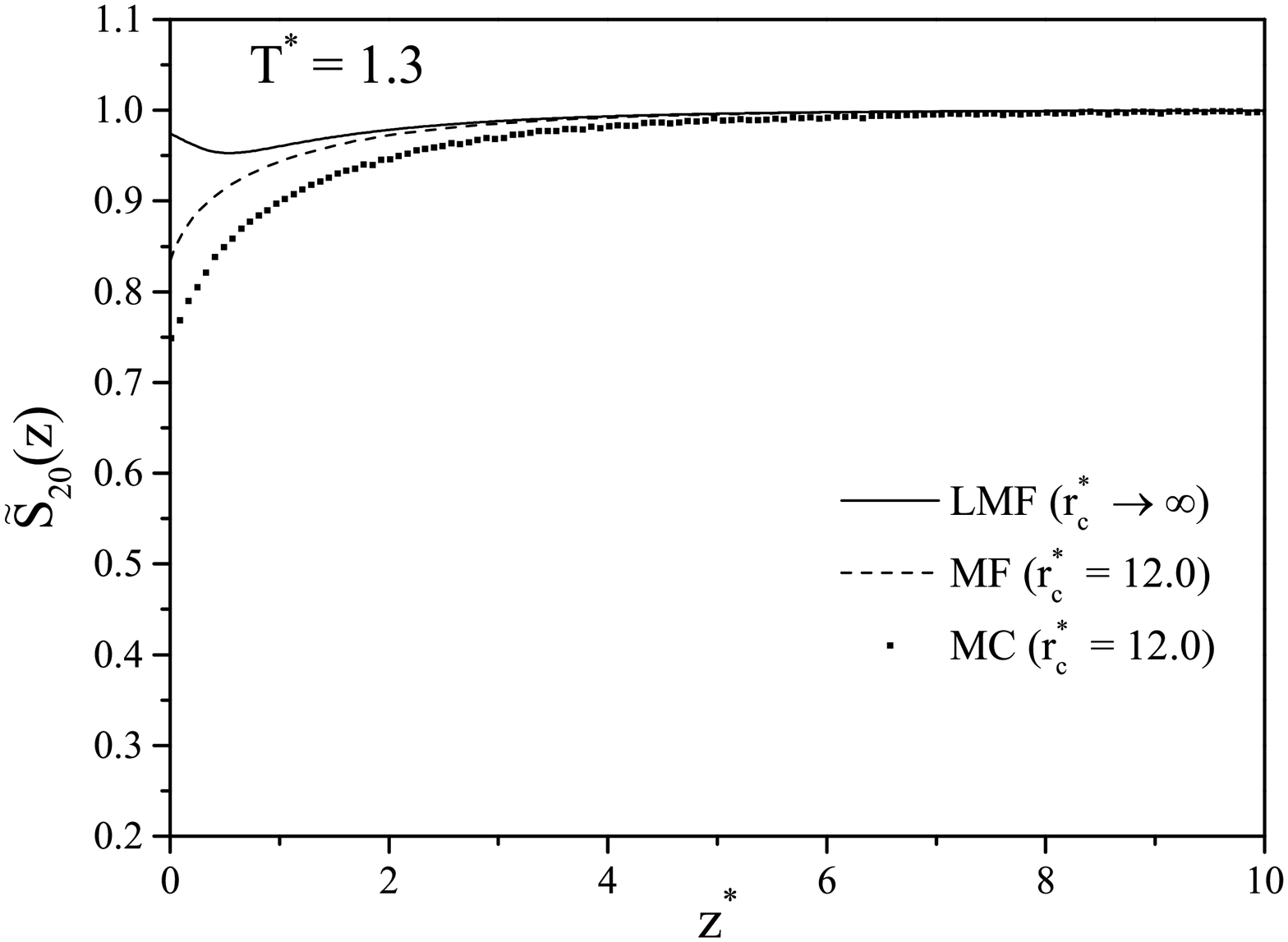}
\includegraphics [width=0.49\textwidth]  {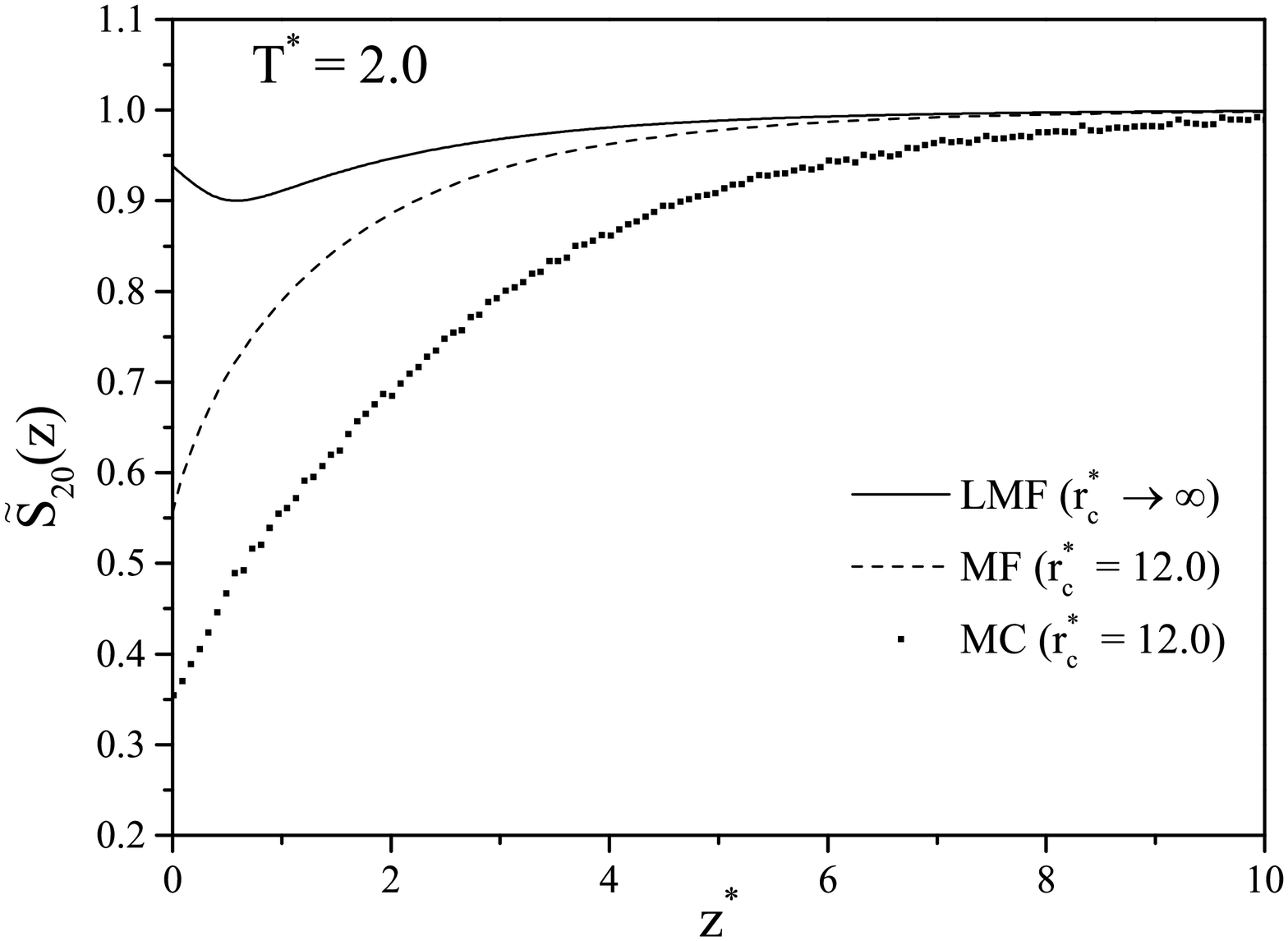}
\caption{Profiles of the normalized order parameter $S^*_{20}(z)/S_\text{b}$ of the MS fluid near a hard wall obtained
in the LMF and MF approximations and with the use of the MC simulation method.
The results were obtained at temperature $T^*=1.3$ (left-hand panel) and $2.0$ (right-hand panel).}
\label{fig:Fig3}
\end{center}
\end{figure}

A different situation is observed for the order parameter {profile}
$S^*_{20}(z)$ of the MS fluid near a hard wall, which is presented in
figure~\ref{fig:Fig3} as a normalized quantity
$\tilde{S}(z)=S^*_{20}(z)/S_\text{b}$. We take such a presentation because of an
essential deviation of the theory from the simulations results for bulk order
parameters (figure~\ref{fig:Fig1}). Using the normalization we can consider
all profiles on the same plot. Nevertheless, even with this normalization, one
can observe an important quantitative and qualitative difference between not
only the theory and simulations, but also between the MF and LMF
approximations (figure~\ref{fig:Fig3}). In the LMF approximation, a minimum
of $\tilde{S}(z)$ is found at $z^*=0.54$ and $z^*=0.58$ at $T^*=1.3$ and
$T^*=2.0$, respectively, while in the MF approximation no minimum has been
noticed at all. At the same time{,} the simulation results do not give any
evidence of a $\tilde{S}(z)$ minimum either. In this context, the MF
results {are similar} to those obtained from simulations. However, all
the considered approaches give completely different contact values of
$\tilde{S}(z)$. The contact value of the order parameter obtained from the
simulation is the lowest one, then {an} essentially larger $\tilde{S}(z)$ is
given by the MF approximation and the largest value of $\tilde{S}(0)$ close
to $1.0$ is calculated from the LMF approximation. In the bulk region (far
enough from the wall), all the profiles converge to $1.0$ as expected. The
only difference is found in the rates of this convergence, which is lower in
the MF approximation than in the LMF approximation, and in the simulations it
is the lowest one. Also, it is worth noting that the rate of convergence of
$S^*_{20}(z)$ to $S_\text{b}$ [i.e., $\tilde{S}(z)$ to 1.0] decreases with the
temperature.

Before discussing the contact values obtained in our study we will try to
understand the behavior of the order parameter profile $S^*_{20}(z)$ and how
it relates to the density profile $\rho^*(z)$. First of all, we have noticed
that the order parameter of fluid particles next to the wall ($z^*\rightarrow
0$) is smaller than the bulk order parameter $S_\text{b}$. However, the density
profiles (figure~\ref{fig:Fig2}) for the same $z^*$ is higher than the bulk
density. From the knowledge of the bulk, for which the higher density leads
to the higher order parameter, one can expect that the order parameter near
the wall is also higher than $S_\text{b}$. Therefore, we encounter a contradiction,
since both the MF approximations and simulation predict the values of
$S^*_{20}(z)$ {smaller} than $S_\text{b}$ everywhere except the bulk-like region
where $S^*_{20}(z)$ is equal to $S_\text{b}$ (figure~\ref{fig:Fig3}). Apparently we
are dealing with two competing effects: densification of fluid particles near
the wall, which should increase $S^*_{20}(z)$ at small $z^*$ and the absence
of fluid particles beyond the wall, which should decrease $S^*_{20}(z)$.
Following the obtained results, one can conclude that the latter effect is
more essential for the system considered in our study.

\subsection{Contact values}

The density and order parameter profiles of a MS fluid near a hard wall
are used to calculate the corresponding contact values $\rho^*(0)$ and $S^*_{20}(0)$ as functions of
the temperature. These contact values can be calculated from the expressions
of the contact theorem (CT) (\ref{eq:CT2b}) and (\ref{eq:CTS1}),
which in the MF approximation reduce to the relations (\ref{contactd}) and (\ref{contacts}),  respectively. In the LMF approximation, the contact
values $\rho^*(0)$ and $S^*_{20}(0)$ satisfy the relations (\ref{contactd}) and (\ref{contacts}) automatically due to the boundary conditions
applied in the solution of the corresponding differential equation. In the MF approximation, the situation
is different for $\rho^*(0)$ and $S^*_{20}(0)$. As it was shown in \cite{HolKrav13}, the contact value for the density profile
$\rho^*(0)$ satisfies the relation (\ref{contactd}), although it has not been proven that the contact value of the order parameter $S^*_{20}(0)$
obtained in the MF approximation should satisfy the relation (\ref{contacts}).

\begin{figure}[!t]
\begin{center}
\includegraphics [width=0.49\textwidth]  {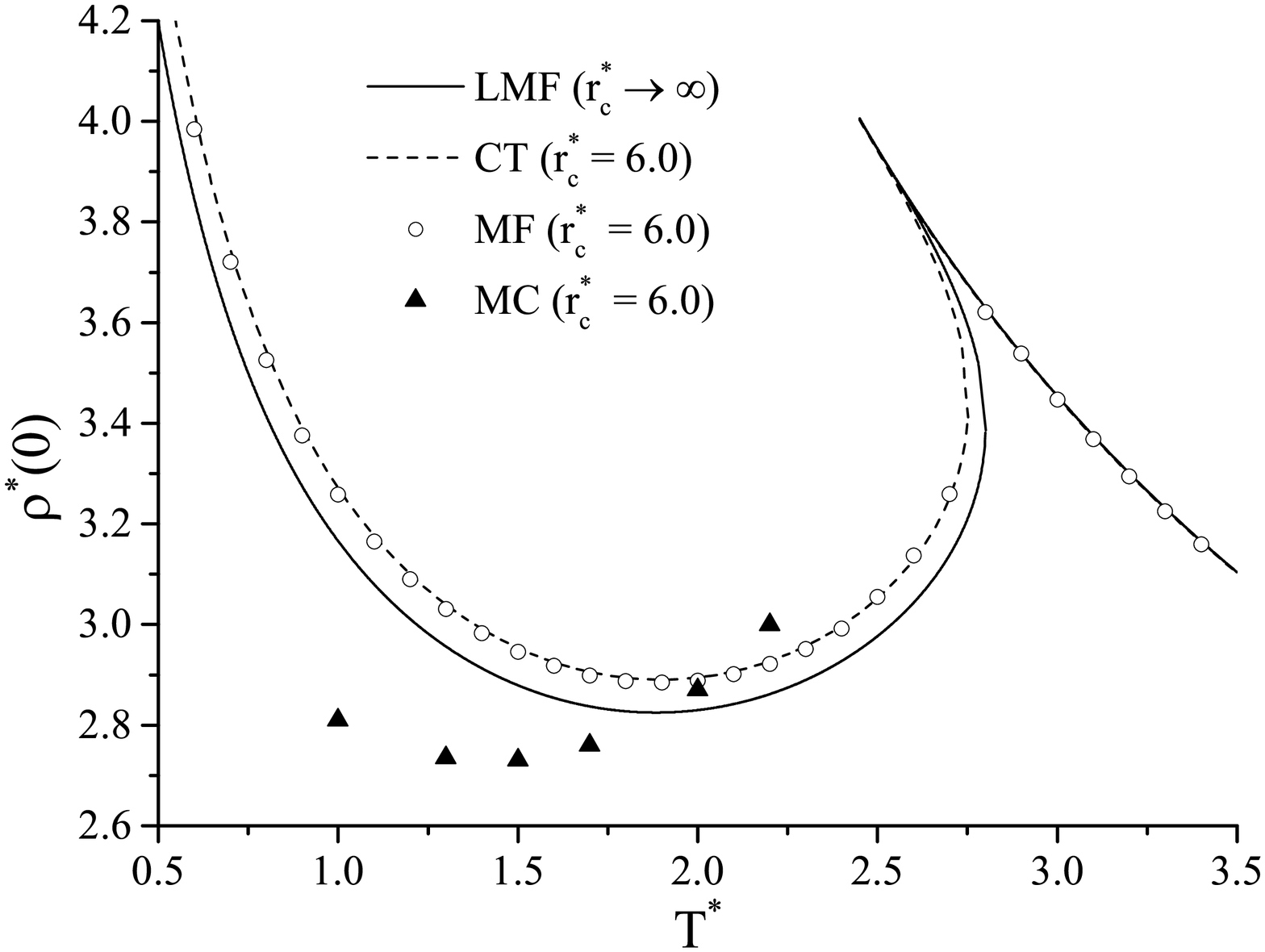}
\includegraphics [width=0.49\textwidth]  {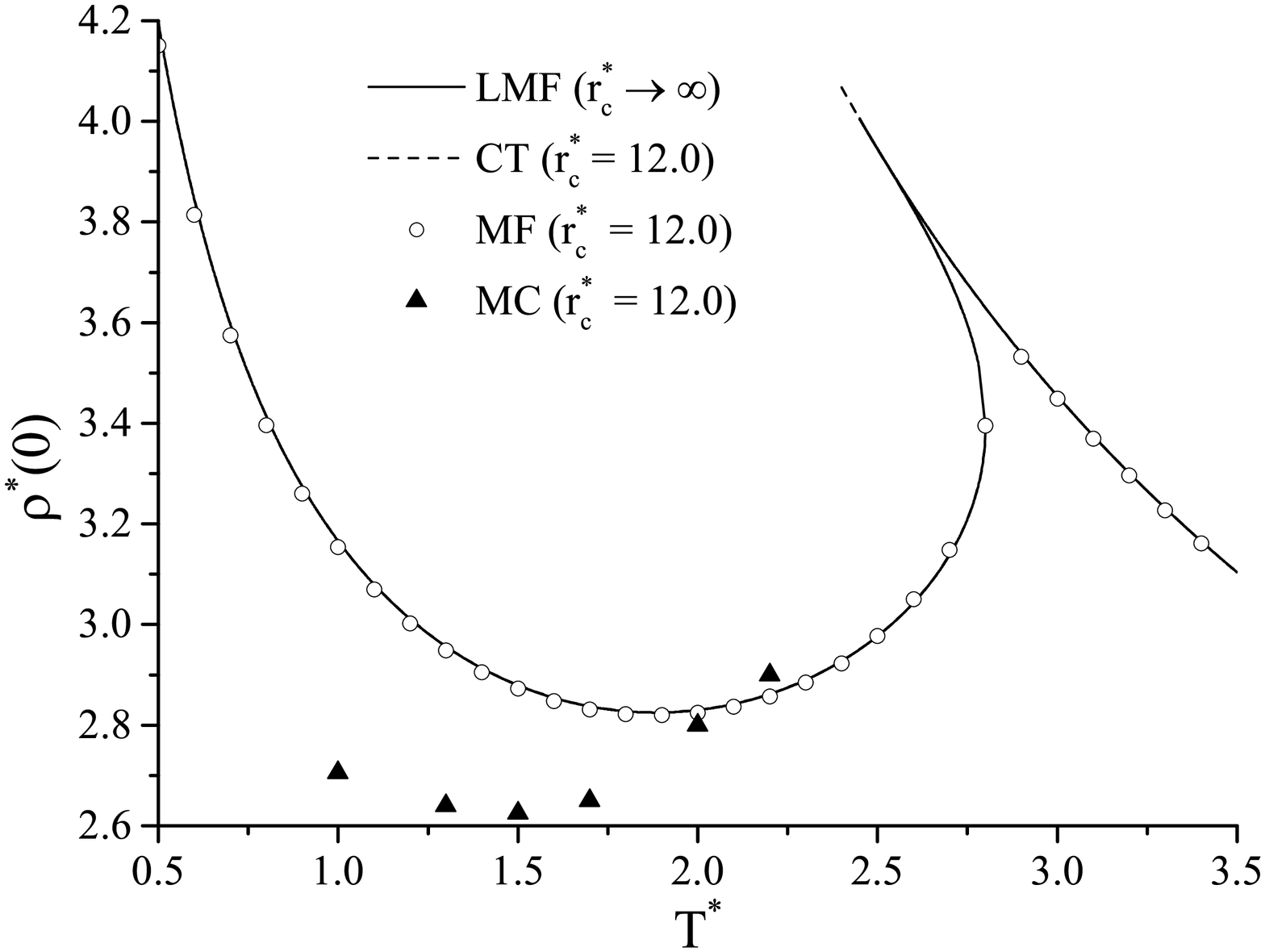}
\caption{Contact value of the density profile as a function of the temperature obtained by different approaches. Two cut-off radii are used:
$r^*_\text{c}=6.0$ (left panel) and $r^*_\text{c}=12.0$ (right-hand panel).}
\label{fig:Fig4}
\end{center}
\end{figure}

First we consider the temperature dependencies of the contact value of
density profiles obtained with the different cut-off radii $r^*_\text{c}=6.0$ and
$12.0$ (figure~\ref{fig:Fig4}). As can be seen, the CT leads to the same
result as the MF approximation. A small difference between the LMF and the MF
approximation (or the CT) appears when the cut-off radius is equal to
$r^*_\text{c}=6.0$ (figure~\ref{fig:Fig4}, left-hand panel). If the cut-off radius is
increased to $r^*_\text{c}=12.0$, equivalent results are obtained in all of the
approximations (figure~\ref{fig:Fig4}, right-hand panel). At the same time, the
agreement between the theoretical approaches and the simulations is mostly
qualitative. We would like to draw attention to the non-monotonous behavior
of the contact value $\rho^*(0)$ with a distinct minimum observed both in the
simulations and in the theoretical predictions. To understand this
interesting effect, we should analyze the temperature dependence of the
density contact value  more in detail.

As it is seen in figure~\ref{fig:Fig4}, at low temperatures, $\rho^*(0)$ is
large and lowers as the temperature increases. This is related to {the}
reduction of the repulsive contribution of the pair
potential~(\ref{potential}). Since in our model a soft repulsive interaction
is used, it becomes weaker if the temperature increases. It should be noted
that the attractive part of the pair potential in our model is rather small
and mainly affects orientational properties of fluid particles. At the same
time, a relative orientation can indirectly strengthen the fluid-fluid
repulsion contribution by reducing the attractive interaction term.
Therefore, at some point in the nematic phase, {when the order parameter
remains sufficiently small,} the repulsion becomes stronger causing an
increase of the contact value $\rho^*(0)$. This effect is observed in
figure~\ref{fig:Fig4} at temperature $T^*=1.885$, where $\rho^*(0)$ reaches
its minimum and starts to increase rapidly until the fluid becomes totally
isotropic [$S^*_{20}(z)=0$]. In the isotropic phase, the fluid particles are
totally orientationally disordered, and a further temperature increase leads
to the weakening of the repulsion. Thus, a continuous decrease of $\rho^*(0)$
is obtained at high temperatures. It should be noted that the explanation
presented here concerns solely, models with a pair potential consisting of a
soft-core term combined with a Maier-Saupe attractive potential. In the case
of a hard-core type of repulsive interaction, the temperature dependence can
be opposite and the effect of orientational ordering can be completely
different.

\begin{figure}[!b]
\begin{center}
\includegraphics [width=0.49\textwidth]  {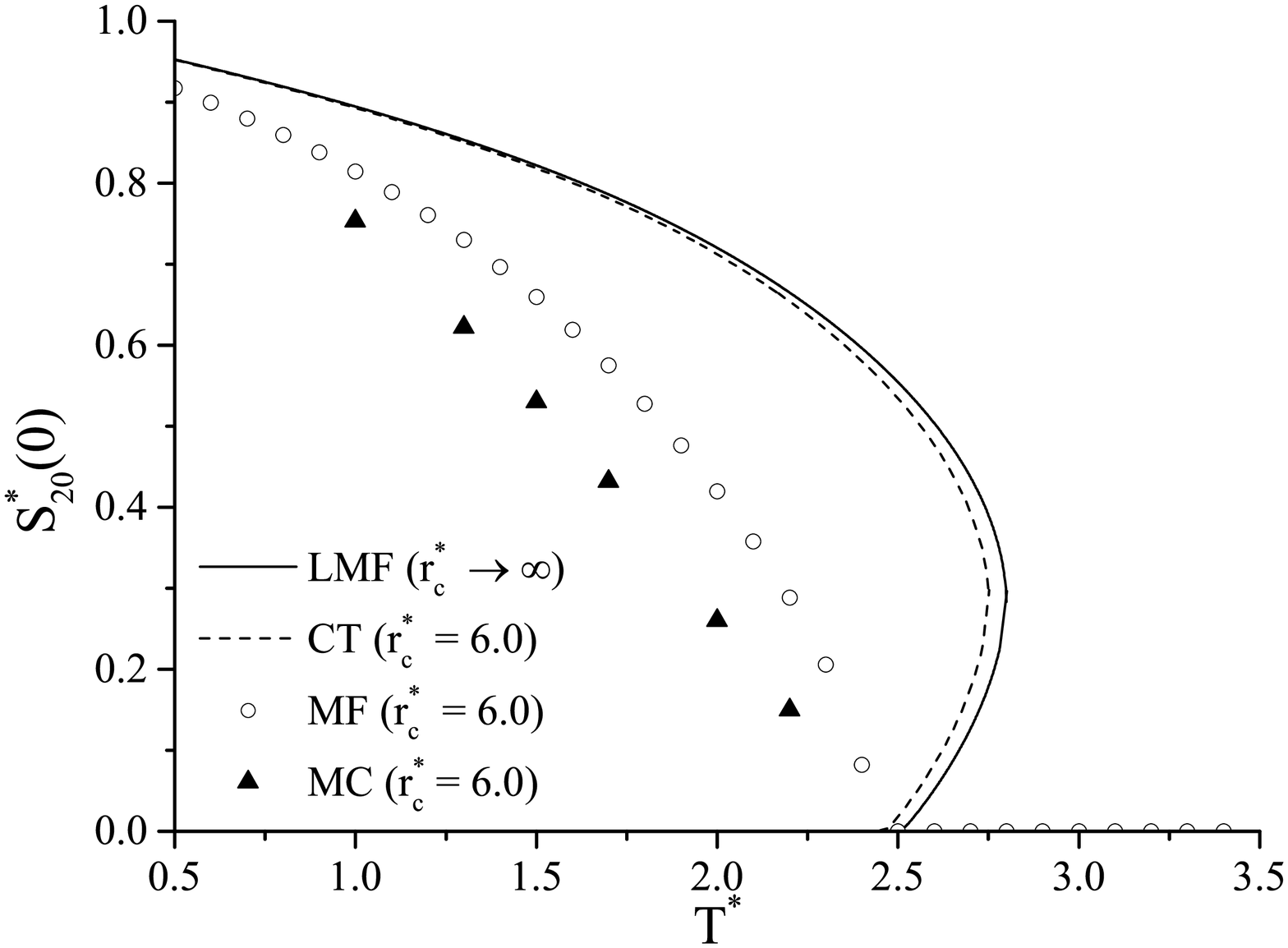}
\includegraphics [width=0.49\textwidth]  {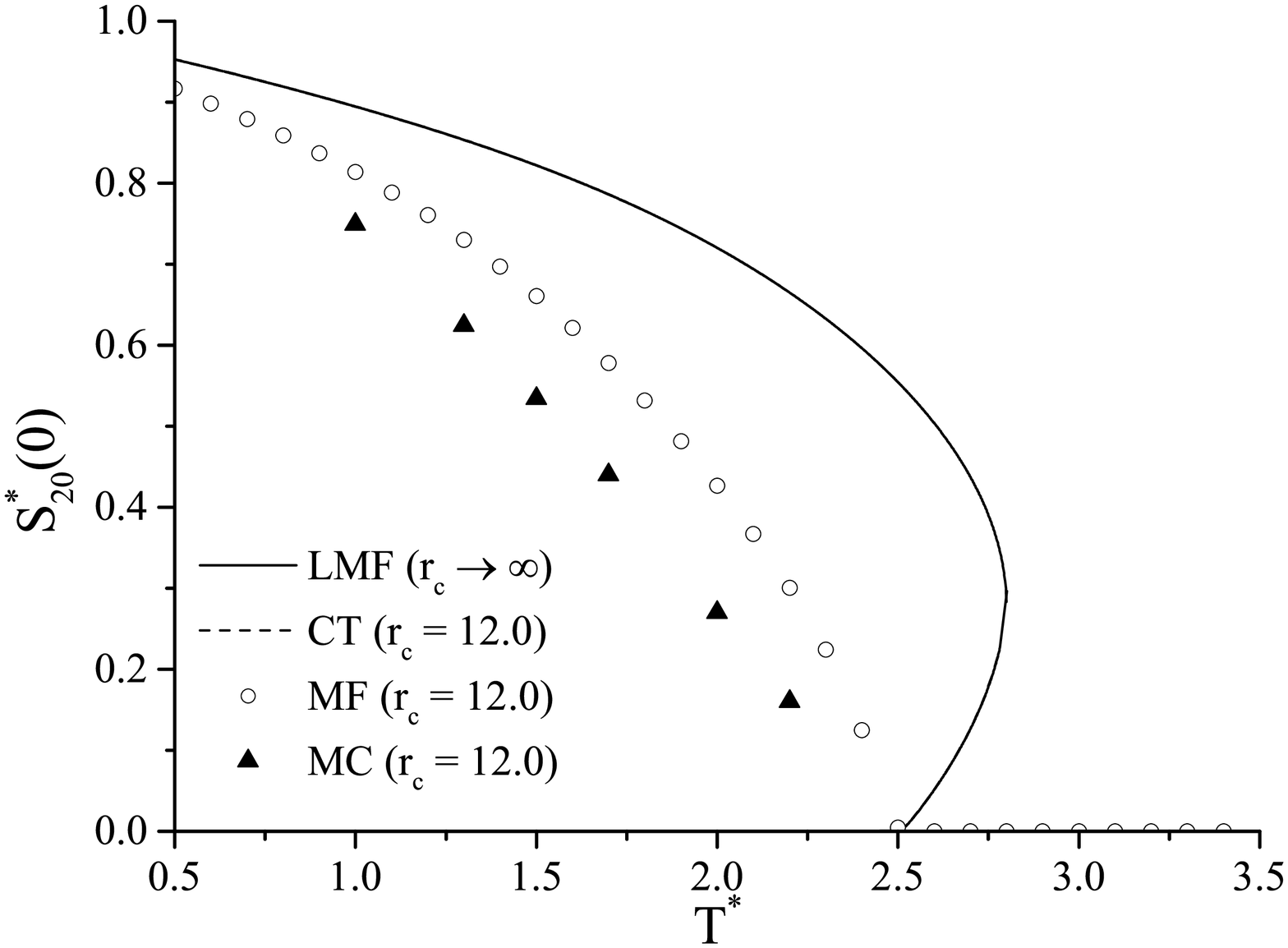}
\caption{Contact value of the order parameter profile as a function of the temperature obtained by different approaches.
Two cut-off radii are used: $r^*_\text{c}=6.0$ (left-hand panel) and $r^*_\text{c}=12.0$ (right-hand panel).}
\label{fig:Fig5}
\end{center}
\end{figure}

While the contact value of the density profile is rather understandable, the
behavior of the contact value of the order parameter profile is not so clear.
First of all, as it has been already shown, there is an essential inconsistency
between the MF and LMF approximations (figure~\ref{fig:Fig3}). There is also
a problem if one compares the contact values $S^*_{20}(0)$ obtained in the MF
and LMF approximations and those calculated from the CT theorem. As can be
seen in figure~\ref{fig:Fig5}, the MF results significantly differ from those
of the LMF and the CT. A perfect agreement of the LMF and the CT appears due
to the definition of the boundary conditions used in the LMF
approximation, which are taken to fit the CT theorem (\ref{contacts}). A
deviation of the CT from the LMF is seen only for the case of $r^*_\text{c}=6.0$
chosen in the CT (figure~\ref{fig:Fig5}, left-hand panel). For $r^*_\text{c}=12.0$, the
contact values of the order parameter of a MS fluid obtained from the CT and
the LMF totally coincide. At the same time, the MF approximation gives much
lower $S^*_{20}(0)$, although it is closer to the simulation results than the
CT and LMF. {We assume that the difference between the CT and the MF
approaches is related to a poor description of the bulk partial pressure
$P(\Omega_{1n})$. In order} to improve the results obtained from the CT (or the
LMF approximation), one should take into account higher order terms, for
instance the Gaussian fluctuations \cite{KravHol13}. Also, it should be noted
that in contrast to the CT and LMF approximation, $S^*_{20}(0)$ obtained from
the MF approximation and the simulations decays to zero before the critical
region is reached. Thus, the convex part of $S^*_{20}(0)$ is absent for them.
It means that in the critical region there is no orientational ordering of
fluid particles at a  contact with the wall.

\section{Conclusions}

Within the framework of the density field theory, we have formulated the mean
field (MF) approximation as a starting point for the theoretical description
of a nematic fluid at a hard wall. Using the developed  approach we have
investigated the density and order parameter profiles of a confined
Maier-Saupe nematogenic fluid with an isotropic Yukawa-like repulsion.
Theoretical predictions have been compared with analytical results obtained
within the framework of the linearized mean field (LMF)
approximation~\cite{HolKrav13}. For the density profile, the results obtained
in the MF and LMF approximations are in qualitative agreement with computer
simulations data. For the order parameter profile, the results of the MF and
LMF approximations have qualitatively different behaviors. In the LMF
approximation, a minimum of the order parameter profile is present, while in
the MF approximation no minimum has been observed. The results of computer
simulations do not give any evidence of the existence of a minimum in the
$z$-dependence of the order parameter profile either. In this context, the MF
and computer simulations results are qualitatively similar. This
contradiction in the description of the density and order parameter profiles
within the framework of the LMF approximation is connected with the problem of
the respective contact theorems in
forms (\ref{contactd}) and (\ref{contacts}) which are used as boundary
conditions in the solution of the corresponding system of differential
equations. As it was shown in reference~\cite{HolKrav13}, the contact value of the
density profile $\rho(0)$ satisfies the relation~(\ref{contactd}) but the
validity of the relation (\ref{contacts}) for the contact value of the order
parameter $S_{20}(0)$ is not evident.
Moreover, the comparison with the results of computer simulations
in figure~\ref{fig:Fig5} shows that relation~(\ref{contacts})  is probably
incorrect. As we can see from figure~\ref{fig:Fig5}, the LMF approximation can
give a better result if we change the boundary condition for the order
parameter profile by correcting the contact value for $S(0)$
obtained within the framework of the MF approximation. We hope that the
considered problem can be better understood by going beyond the MF
approximation and including a contribution from fluctuations.

The temperature dependencies of the contact values of the density and order
parameter profiles have been analyzed more in  detail. We have found a
non-monotonous behavior of the density contact value $\rho(0)$ as a function
of the temperature with a distinct minimum observed in both theoretical
predictions and computer simulations. This non-monotonous temperature
dependence of $\rho(0)$ is explained by the competition of the soft isotropic
repulsive and the soft anisotropic attractive contributions. For the contact
value of the order parameter, we have observed a monotonous decrease with
an increase of temperature. Both the simulations and the MF results indicate that
there is no orientational ordering of fluid particles at the contact with the
wall when the fluid is in the critical region.

The results presented in this paper have been obtained within the framework of
the MF approximation. We note that the agreement between theoretical
predictions and computer simulation data is mostly qualitative. For a better
theoretical description one should take fluctuations into account. For the
bulk properties of the model under consideration, the influence of the
contribution from fluctuations was already discussed in
reference~\cite{KravHol13}. It was shown that the singlet distribution function
reduces to the form (\ref{singlet_bulk}) with a change of~$\kappa_{2}^{2}$
to~$\kappa_{2}^{2}\,t$, where $t=1-\beta A_{0}\alpha_{2}^{2}/(\lambda_{0}+\lambda_{2})$.
It was shown that the temperature trend of deviation between the order
parameter $S_\text{b}$ calculated with fluctuations included and the MF value is
the same as that between computer simulations and the MF value presented
in figure~\ref{fig:Fig1}. In the next paper we plan to include the
contribution from fluctuations in the description of a MS nematogenic fluid
at a hard wall similar to the way it was done for isotropic Yukawa
fluids~\cite{DdiCaprio11,KravPat15}.

\ukrainianpart

\title{Нематогенний плин Майєра-Заупе з ізотропним юкавівським відштовхуванням біля твердої поверхні: наближення середнього поля}
\author{М. Головко\refaddr{label1}, Т. Пацаган\refaddr{label1}, І. Кравців\refaddr{label1}, Д. ді Капріо\refaddr{label2}}
\addresses{
\addr{label1} Інститут фізики конденсованих систем НАН України,
вул. І.~Свєнціцького, 1, 79011 Львів, Україна
\addr{label2} Лабораторія електрохімії, хімії поверхонь і енергетичного моделювання, відділення хімії вищої національної школи ПаріТех,
вул. П. і М. Кюрі, 11, 75005 Париж, Франція
}

\makeukrtitle

\begin{abstract}
\tolerance=3000%
В рамках теорії поля густини сформульовано наближення середнього поля для дослідження властивостей нематогенного плину Майєра-Заупе біля твердої поверхні. У лінеаризованому наближенні середнього поля розраховано аналітичні вирази для профілю густини та профілю параметра порядку. Детально проаналізовано залежність контактних значень профілів густини та параметра порядку від температури. Для оцінки застосовності використаних наближень проведено порівняння отриманих теоретичних результатів з оригінальними даними комп'ютерного моделювання.
\keywords нематогенний плин Майєра-Заупе, теорія поля, поверхня, контактна теорема, потенціал Юкави
\end{abstract}

\end{document}